# Competing Chern states revealed by quasiparticle charging in moiré rhombohedral graphene

Hongyuan Li[1,2†*], Zuhan Geng[1,3†], Junseok Seo[4†], Chenxi Xu[1], Yifan Jiang[1], Shenyong Ye[4], Zhenqi Hua[4,5], Jiabin Xie[1], Lujin Min[6], Kenji Watanabe[7], Takashi Taniguchi[8], Kenji Yasuda[6], Xiaomeng Liu[1], Long Ju[4], Jie Shan[1-3,6*], and Kin Fai Mak[1-3,6*]

[1]Laboratory of Atomic and Solid State Physics, Cornell University, Ithaca, NY, USA
[2]Kavli Institute at Cornell for Nanoscale Science, Ithaca, NY, USA
[3]Max Planck Institute for the Structure and Dynamics of Matter, Hamburg, Germany
[4]Department of Physics, Massachusetts Institute of Technology, Cambridge, MA, USA
[5]Department of Physics, Florida State University, Tallahassee, FL 32306
[6]School of Applied and Engineering Physics, Cornell University, Ithaca, NY, USA
[7]Research Center for Electronic and Optical Materials, National Institute for Materials Science, Tsukuba, Japan
[8]Research Center for Materials Nanoarchitectonics, National Institute for Materials Science, National Institute for Materials Science, Tsukuba, Japan

[†]These authors contributed equally to this work.
[*]Emails: hl2483@cornell.edu, jie.shan@mpsd.mpg.de, kinfai.mak@mpsd.mpg.de

**Moiré materials realize a versatile platform for exploring the physics of fractional Chern insulators (FCIs)[1-21]. The recently observed evolution from FCIs to an extended quantum anomalous Hall background upon lowering the electronic temperature in moiré rhombohedral graphene (mRG)[8] raises a fundamental question: Is it caused by a failure to equilibrate the edge states of an FCI or by a genuine phase transition in the bulk from an FCI to a generalized anomalous Hall crystal? Here we address this question by probing quasiparticle charging in a mesoscopic mRG antidot device and by bulk resistance measurements, both of which are bulk-sensitive and free from complications from edge states. Tunneling to the mRG antidot reveals quasiparticles carrying one electron charge for both Chern states at filling factors $\nu = 1$ and 2/3 at low temperatures. Temperature dependence measurements of the bulk resistance near $\nu = 2/3$ further suggest a thermodynamic phase transition from an FCI to a generalized anomalous Hall crystal at temperatures below about 150mK. The results clearly exclude the edge state equilibration scenario and favor the phase transition scenario. Our work establishes mesoscopic probes as a powerful approach to uncover competing ground states in moiré materials and provides a basis for probing fractionalized excitations in FCIs.**

## Main

Moiré materials provide a versatile platform to engineer electronic correlations and band topology, whose interplay gives rise to new states of matter. A prominent example is the observation of FCIs in both moiré semiconductors[1-5] and moiré rhombohedral graphene[6-10,21]. Whereas FCIs are realized in experimental systems involving flat Chern bands, the necessary ingredients for stabilizing a FCI (beyond an electron band with sizeable Berry curvatures and strong interactions) remain unclear[11,15,22,23]. For instance, neither Chern bands[22] nor ideal quantum geometry[11] is required. A better understanding of this question and of the relevant competing phases for FCIs is thus essential for further exploration of FCI physics and for potential applications of FCIs.

Moiré rhombohedral graphene (mRG) has attracted particular attention in this context. Specifically, recent experiments had shown that a fractional quantum anomalous Hall (QAH) effect can be overtaken by an extended (integer) QAH background upon lowering the electronic temperature of the material[8,24,25] (A related reentrant QAH effect was also reported in twisted bilayer $MoTe_2$[26]). In fractional quantum Hall states (a close relative of FCIs), deviations of the Hall response from fractional quantization had been observed; a plausible explanation is a failure of edge state equilibration in micron-scale device channels rather than a change in the nature of the bulk state[27-29]. For instance, the upstream and downstream edge modes of a filling 2/3 fractional quantum Hall state[30-33] may not be able to fully equilibrate without sufficient hybridization[27-29,32,33], leading to a Hall resistance $R_{xy}$ less than $\frac{3}{2}\frac{h}{e^2}$ (and approaching $\frac{h}{e^2}$ if the integer downstream mode dominates). (Here $h$ and $e$ denote the Planck's constant and the electron charge, respectively.) An important question then arises: Is the extended QAH background in mRG connected to edge state equilibration or to a genuine phase transition in the bulk involving competing ground states?

While this question may not be resolved by conventional Hall bar measurements that are mainly edge-sensitive, we address this question here by bulk-sensitive measurements performed on a gate-defined antidot device of rhombohedral pentalayer graphene angle-aligned with hexagonal boron nitride (hBN). Charging to the antidot reveals quasiparticles carrying one electron charge for both Chern states at $\nu = 1$ and 2/3 at low temperatures. The result clearly rules out the edge state equilibration scenario. Measurements of the bulk resistance near $\nu = 2/3$ further support the presence of a thermodynamic phase transition from a FCI to a generalized anomalous Hall crystal at temperatures $T \lesssim 150$mK. The result favors the competing ground state scenario.

**mRG antidot devices**

Figure 1a illustrates the device structure used in this study. Pentalayer rhombohedral graphene is encapsulated between hBN dielectrics and angle-aligned to the bottom hBN layer (Fig. 1b), forming a moiré lattice with period $a_M \approx 12.7$nm (see Extended Data Fig. 2 for calibrations). To stabilize the rhombohedral polytype relative to its Bernal counterpart and to enhance the fabrication yield, we developed a method by adsorbing $NO_2$ molecules onto the surface of the rhombohedral graphene to hole-dope the material, which is known to stabilize the polytype against structural relaxation by lowering its free energy relative to the Bernal polytype[34]. The encapsulated stack is sandwiched between top and bottom graphite gate electrodes to enable independent control of the filling factor $\nu$ and the vertical displacement field $D$ in the overlapped regions of the two gates. The filling factor is defined as $\nu \equiv n/n_M$, where $n$ is the doping density and $n_M = \frac{2}{\sqrt{3}(a_M)^2} \approx 0.73 \times 10^{12}$cm$^{-2}$ is the moiré density. (See Methods for details on device fabrication and doping density calibration.)

To define an antidot in the mRG by gating, we first used a reported AFM lithographic technique[35] to etch a hole and, right next to it, two "fingers" in the top graphite gate electrode (AFM denotes atomic force microscope)[36]; the technique avoids residual contamination from electron-beam lithography and enables a clean mesoscopic structure, as illustrated in Fig. 1c. We then transferred an additional control gate (made of hBN and graphite) on top of the prepatterned graphite top gate.

In our experiment, the top and bottom gates combined can set the mRG sample to a Chern insulating state in the overlapped regions of the gates; the control gate can define, through the etched regions of the top gate, an antidot and two electrodes forming tunnel contacts to the antidot (red regions in Fig. 1a). The electron number in the antidot is controlled by fine-tuning the voltage, $V_{CG}$, applied to the control gate; coarse-tuning of $V_{CG}$ pinches the quantum point contact (QPC) between the antidot and the tunneling electrode and thus modifies the tunneling barrier height to the antidot (See Methods and Extended Data Fig. 3 for details).

We finished the device in a Hall bar geometry (Fig. 1d and optical image in Extended Data Fig. 1) for both regular Hall bar measurements of the sample resistivities and tunneling measurements on the antidot. Quasiparticle tunneling to the antidot was monitored by measuring the diagonal resistance[37-40], $R_D$, which is defined as the ratio of the diagonal voltage drop across the antidot to the source-drain current (Fig. 1d). We focus on the weak tunneling regime[41-43], in which the filling factor in the QPC constriction and the bulk channel are nearly the same (see Extended Data Fig. 3 for supporting evidence). For an integer Chern insulator with Chern number $C = 1$, $R_D$ is close to the quantized value $\frac{h}{e^2}$ in the weak tunneling limit; small deviations from $R_D = \frac{h}{e^2}$ are caused by quasiparticle tunneling from the tunneling electrodes to the antidot[37]. We will also focus on the regime, in which the antidot and the bulk channel share the same charge carrier type (n-type). (See Methods and Extended Data Fig. 4 for similar results in the regime of opposite carrier types.) Although the antidot is not set to charge-neutrality in our experiment, its total charge is quantized in units of the quasiparticle charge determined by the nature of its surrounding bulk[44], specifically, integer (fractional) quasiparticle charge for integer (fractional) Chern insulators. The tunneling probe of antidot charging is bulk-sensitive.

We first examine the four-terminal longitudinal ($R_{xx}$) and Hall ($R_{xy}$) resistances of the bulk channel (see Fig. 1d) as a function of $\nu$ and $D$ at $T = 400\,\text{mK}$ (Fig. 1e,f). $R_{xx}$ and $R_{xy}$ are symmetrized and anti-symmetrized, respectively, under a perpendicular magnetic field $B = \pm 50\text{mT}$. We focus on high displacement fields and fillings in between 0.3 and 1.1, where the integer and fractional Chern states emerge. A QAH effect with $R_{xx} \approx 0\,\Omega$ and $R_{xy} = \frac{h}{e^2}$ is observed near the fully developed $\nu = 1$ Chern insulating state. $R_{xx}$ dips and $R_{xy}$ peaks are also observed at $\nu = 1/3$, 2/5, 3/5 and 2/3 for the partially developed FCIs. (The emergence of FCIs is further confirmed by the magnetic field dispersion of the states in Extended Data Fig. 5.) Moreover, a non-topological insulating state with large $R_{xx}$ appears at low displacement fields and filling factors (i.e. the top left corner of the phase diagram). The results are consistent with Ref. [6].

Upon lowering the (lattice) temperature to $T = 12\text{mK}$, the FCI features largely disappear in the $R_{xx}$ and $R_{xy}$ maps (Fig. 1g,h); the $R_{xx}$ dips and $R_{xy}$ peaks are replaced by an extended region of small $R_{xx}$ and nearly quantized $R_{xy} = \frac{h}{e^2}$. The temperature evolution for $R_{xx}$ and $R_{xy}$ at a constant displacement field line cut ($D \approx -0.93\text{V/nm}$) are shown in Fig. 1i and 1j, respectively. For $\nu \gtrsim 0.5$, $R_{xx}$ decreases with decreasing temperature and the $R_{xx}$ dip at $\nu = 2/3$ evolves into a small $R_{xx}$ peak at low temperatures. Meanwhile, the $R_{xy}$ peaks corresponding to the partially developed FCIs now converge to a nearly quantized $R_{xy} = \frac{h}{e^2}$ plateau spanning the range $0.5 \lesssim \nu \lesssim 1.0$. The results are consistent with the reported emergence of an extended QAH background at low temperatures[8]. Note that the imperfect fractional quantization of $R_{xy} = \frac{h}{\nu e^2}$ at $\nu = 3/5$ and 2/3 and at elevated temperatures is likely caused by the combined effects of thermal excitations in the bulk for $T \gtrsim 300$mK and the presence of competing ground states for $T \lesssim 100$mK (see below). The degree of quantization is also device dependent (likely due to variations in the twist angle), with some devices showing more robust quantization (see Ref. [8] and Extended Data Fig. 9).

**Quasiparticle charging near $\nu = 1$**

Next, we examine quasiparticle charging to the antidot near $\nu = 1$. Figure 2a shows the diagonal resistance $R_D$ as a function of the control gate $V_{CG}$ at $B = 50$mT and $T = 12$mK; we kept the bulk channel at $\nu = 0.94$ and $D = -0.94$V/nm, i.e. near the integer Chern insulator. (Note that $V_{CG}$ is negative because $V_{BG}$ is positive.) Periodic peaks in $R_D$ on top of a nearly quantized background at $\frac{h}{e^2}$ are observed, consistent with quasiparticle charging to the antidot probed in the weak tunneling regime as mentioned above. Quasiparticle charging to the antidot is further evidenced in Fig. 2b by the Coulomb diamond pattern observed in the map of $R_D$ versus $V_{CG}$ and $V_{DC}$ (the DC source-drain bias voltage). By varying $V_{CG}$, quasiparticles are injected into and bound with the antidot; each additional quasiparticle induces a shift in the antidot chemical potential by one discrete quantized energy level, producing a resonance tunneling event that increases $R_D$ (Ref. [37]). We can estimate a quantized energy level separation on the order of 100μeV from the Coulomb diamond.

We also study the dependence of quasiparticle charging on $\nu$ and $D$ along traces i and ii shown in Fig. 1g. To this end, we Fourier transformed $R_D$ (at $V_{DC} = 0$V) with respect to $V_{CG}$ at each $\nu$ and $D$ and show the dependence of the Fourier transformed spectrum on $D$ and $\nu$ in Fig. 2c and 2d, respectively. A Fourier peak with frequency $f_{CG} \approx 100\,\text{V}^{-1}$ (corresponding to the Coulomb oscillations shown in Fig. 2a) is observed near $\nu = 1$ and over the entire range of the displacement field for the $\nu = 1$ Chern insulator. The results confirm quasiparticle charging to an antidot embedded in the bulk of a Chern insulator, rather than charging to parasitic quantum dots unintentionally formed in the device.

This point is further illustrated by the observation of Laughlin charge pumping in Fig. 2e, which shows $R_D$ as a function of $V_{CG}$ and $B$ (at $\nu = 0.98$ and $D = -0.95$V/nm). Periodic stripes in the diagonal direction with a positive slope are observed. The $V_{CG}$ period ($\Delta V_{CG} \approx 9.5$mV) at fixed $B$ corresponds to quasiparticle charging by the control gate as discussed above. The magnetic field period ($\Delta B \approx 55\,$mT) at fixed $V_{CG}$ corresponds to Laughlin charge pumping[45]: Threading a magnetic flux quantum $\frac{h}{e}$ through an antidot embedded in the bulk of a $C = 1$ Chern insulator transfers a quasiparticle (of charge $-e$) from the antidot to the outer edge of the sample. An increase in $V_{CG}$ is thus required to keep a constant total charge in the antidot, corresponding to the stripes with a positive slope in Fig. 2e.

In the "Coulomb-dominated" regime[41,46,47] as in our experiment, the field period is $\Delta B = \frac{h}{eA}$ for a $C = 1$ Chern insulator ($A$ is the antidot area); the $V_{CG}$ period $\Delta V_{CG} = \frac{e^*}{C_{CG}}$ is proportional to the quasiparticle charge $e^*$ and inversely proportional to the effective control gate capacitance $C_{CG}$ for the antidot. For the $\nu = 1$ Chern insulator, we have $e^* = e$, which can also be confirmed by electrostatics simulations using the actual device geometry and a voltage period $\Delta V_{CG} \approx 9.5$mV (Methods and Extended Data Fig. 7). The field period $\Delta B \approx 55$mT determines an effective antidot diameter of about 310nm. Comparison with the diameter (≈210nm) of the etched hole in the top graphite gate electrode gives a depletion length about 50nm for the antidot. The finite depletion length is likely caused by electric field leakage near the antidot edge and/or by wavefunction delocalization beyond the antidot confinement potential profile.

**Quasiparticle charging near $\nu = 2/3$**

We now turn to quasiparticle charging to the antidot near $\nu = 2/3$. Similar to the case of $\nu = 1$, we observe in Fig. 3a Coulomb oscillations in $R_D$ as a function of $V_{CG}$ on top of a nearly quantized $R_D \approx \frac{h}{e^2}$ background (measured at $\nu = 0.67$, $D = -0.93$V/nm, $B = 50$mT and $T = 12$mK). The voltage period $\Delta V_{CG} \approx 9.7$mV is nearly identical to that near $\nu = 1$. The doping dependence of the Fourier transformed spectrum (with respect to $V_{CG}$) along trace iii in Fig. 1g shows a similar Fourier peak with frequency $f_{CG} \approx 100$V$^{-1}$ and a maximum amplitude near $\nu = 2/3$ (Fig. 3b and 3c). A weak second harmonic peak near $f_{CG} \approx 200$ V$^{-1}$, reflecting non-sinusoidal Coulomb oscillations, is also observed. We also observed Laughlin charge pumping near $\nu = 2/3$ in the map of $R_D$ versus $V_{CG}$ and $B$ (Fig. 3d); a magnetic field period $\Delta B \approx 52$mT nearly the same as that for $\nu = 1$ is obtained. Moreover, the temperature dependence of the Fourier transformed spectrum at $\nu = 2/3$ shows the disappearance of the fundamental Fourier peak at temperatures above ≈100mK (Fig. 3e). The disappearance of Coulomb oscillations at $T \gtrsim 100$mK is likely caused by thermal broadening of the antidot energy levels and/or by thermal activated transport in the QPC constriction that overwhelms the tunneling signature of quasiparticle charging.

In the "Coulomb-dominated" regime[37,38,40,47-49], the quasiparticle charge $e^* = C_{CG}\Delta V_{CG}$ is determined by the voltage period $\Delta V_{CG}$. Assuming a doping-independent gate capacitance $C_{CG}$, which is often an excellent approximation[37,38,40,48], the nearly identical $\Delta V_{CG}$ for $\nu = 1$ and 2/3 implies the same quasiparticle charge $e^* = e$ for both $\nu = 1$ and 2/3. In other words, the $\nu = 2/3$ state is a $C = 1$ integer Chern insulator at temperatures below ≈100mK; the observed extended QAH background near $\nu = 2/3$ is thus not an effect of edge state non-equilibration but a genuine phase transition from an FCI to an integer Chern insulator.

**Competing ground states probed by bulk resistance**

We further examine the phase transition by performing temperature dependence studies on the bulk resistance ($R_{bulk}$) of the sample. This is achieved by grounding electrodes downstream the source electrode and measure the current only out of the electrode immediately upstream the source electrode (Fig. 4a); the two-terminal resistance measured in this geometry provides a proxy for $R_{bulk}$ when the sample is in an insulating state[50-52]. Figure 4b shows $R_{bulk}$ as a function of $\nu$ and $D$ under the same conditions as in Fig. 1g for $R_{xx}$. Insulating behavior with $R_{bulk} \gtrsim 10\frac{h}{e^2}$ is observed over much of the extended QAH and the non-topological insulating regions of the phase space. In particular, the regions near both $\nu = 1$ and 2/3 show very large $R_{bulk}$, which correlates with the nearly vanishing $R_{xx}$ in Fig. 1g. A local $R_{bulk}$ minimum centered at $\nu = 2/3$ is also observed.

We focus on the behavior near $\nu = 2/3$. Figures 4c and 4d show the temperature dependence of $R_{bulk}$ versus $\nu$ along the dashed line in Fig. 4b. As temperature decreases, a sharp resistance peak centered at $\nu = 2/3$ evolves into a dip at $T \lesssim 200$ mK; a highly insulating background surrounding $\nu = 2/3$ appears at $T \lesssim 100$mK. A non-monotonic temperature dependence of $R_{bulk}$ at $\nu = 2/3$ is also observed (Fig. 4e); $R_{bulk}$ reaches the lowest value near $T = 150$mK.

The results suggest the presence of a thermodynamic phase transition at $\nu = 2/3$ with a critical temperature around $T = 150$mK. A schematic phase diagram is shown in Fig. 4f. The high-temperature state is an FCI, as supported by $R_{xx}$ and $R_{xy}$ measurements in Fig. 1 and by the dispersion of the insulating state under a magnetic field in Extended Data Fig. 5 and 6. The low-temperature state is a $C = 1$ integer Chern insulator, as shown by quasiparticle charging in Fig. 3 and also by the magnetic field dispersion. The appearance of a highly insulating background in the immediate vicinity of $\nu = 2/3$ below $T \approx 100$mK shows sensitivity to lattice commensurability; it suggests that charge localization in the moiré lattice driven by the extended-range Coulomb interaction is the driving force for the phase transition[53-56]. (The same interaction is also responsible for the emergence of Wigner-Mott insulators at commensurate fractional fillings in other moiré materials.) A reasonable ground state candidate is thus a generalized anomalous Hall crystal[57-60] carrying a chiral edge state and electrons in the bulk self-organized into a honeycomb lattice (Fig. 4g). Future scanning probe experiments may verify this picture[61].

Finally, we note that the non-monotonic temperature dependence exhibiting a minimum in $R_{bulk}$ near the critical temperature (Fig. 4e) suggests a gap closure at the (possibly continuous) topological phase transition. The survival of the $R_{bulk}$ dip at $\nu = 2/3$ down to the lowest temperatures suggests the presence of residual phase competition between the FCI and the generalized anomalous Hall crystal, which leads to a smaller transport gap exactly at $\nu = 2/3$ compared to its surroundings. Our work establishes mesoscopic transport as a powerful probe for the competing ground states in moiré materials. Future Coulomb screening engineering by tuning the sample-gate distance may fully stabilize the FCI down to the lowest temperatures[15,23] for probing fractionalized excitations using the antidot device architecture[37,39,40,48].

## Methods

### Device fabrication

Thin flakes of graphene/graphite and hexagonal boron nitride (hBN) were mechanically exfoliated from their bulk crystals onto $SiO_2$/Si substrates. The rhombohedral domains of pentalayer graphene were first identified through fast screening with an infrared camera[8] under LED illumination with a center wavelength of 1.6um, followed by Raman mapping to determine the layer stacking order and domain boundary with sub-micron spatial resolution[62]. The rhombohedral domains were then isolated by AFM cutting through the electrode-free local anodic oxidization technique[35]. We also used the same cutting technique to define the antidot and the tunneling electrodes on the top graphite gate.

The fabrication of the complete device stack was separated into two parts: 1) the lower stack of rhombohedral graphene and the bottom gate, and 2) the upper stack of control gate and top gate. Both parts were assembled by the layer-by-layer transfer technique using a polycarbonate thin film on top of a polydimethylsiloxane stamp[63]. The pickup order for the lower stack is hBN, rhombohedral graphene, hBN, and bottom graphite gate; that for the upper stack is hBN, control gate graphite, hBN, AFM-patterned top graphite gate. The lower stack was first released onto a $SiO_2/Si$ substrate with pre-patterned electrodes and cleaned by an AFM in contact mode. The upper stack was then released onto the lower stack. The finished stack was cleaned by the AFM once more. When making the lower stack, $NO_2$ gas (about 1% concentration mixed with air) was adsorbed onto the rhombohedral graphene surface (immersed in $NO_2$ for about 3 minutes) before being picked up and encapsulated by hBN. The $NO_2$ adsorption strongly hole-dopes the graphene and stabilizes the rhombohedral stacking relative to its Bernal counterpart[34]. The $NO_2$ adsorbate trapped in the graphene/hBN interface was then removed by heating the stack to above $200^0$C during the final release process.

The completed stack was then etched into a Hall bar geometry using standard electron-beam lithography and reactive-ion etching. We deposited Cr/Pd/Au (5nm/15nm/50nm) as electrical contacts to graphene/graphite[63]. Three devices were examined in this study. See Extended Data Fig. 8 and 9 for selected results from another two devices.

**Electrical measurements**
Transport measurements on devices 1-3 were performed in a Bluefors LD250 dilution refrigerator at Cornell. A silver-epoxy filter (Basel Precision Instrument MFT25) integrated with a 2-pole RC filter were installed on the mixing chamber plate to filter out the microwave frequency noise and to facilitate electron thermalization. Additional home-made silver-epoxy filters and 2-pole RC filter ($R = 2$kΩ, $C = 15$nF) were installed in the puck right before the sample holder to further reduce noise and for better thermalization. In our measurements, we connected a 20MΩ bias resistor in series with the device to limit the excitation current (<0.5nA). Voltage pre-amplifiers with large input impedance (100MΩ) were used to measure voltage drops in the device. Low-frequency (<5Hz) lock-in techniques were adapted to measure the four-terminal resistances. The voltage drop at the probe electrodes and the source–drain current were recorded simultaneously. Yokogawa GS200 DC voltage sources were used to apply the gate voltages $V_{TG}$, $V_{BG}$ and $V_{CG}$. We also applied a 60V gate voltage using the silicon gate (via Keithley 2400) to lower the mRG contact resistance.

Device 3 was also measured in a Bluefors LD250 dilution refrigerator at MIT (Extended Data Fig. 9). We thermalized twisted phosphorus bronze wires at each stage of the fridge: The wires went through two thermal meanders (one on the mixing chamber plate and the other on the cold finger), ceramic radiofrequency filters (DigiKey), four-stage RC filters, and silver epoxy filters (Basel Precision Instruments MFT25 and home-made one). The setup allows efficient electron cooling. Lock-in techniques with low AC frequency were used to measure the resistance. Voltage preamplifiers (Basel Precision Instruments SP1004) were used to measure voltage drops. Keithley 2400 voltage sources were used to apply all the gate voltages.

**Coarse control by $V_{CG}$**
In Extended Data Fig. 3, we present $R_{xx}$, $R_{xy}$ and $R_D$ as a function of $V_{CG}$ and $\nu$ at $D = -0.83$V/nm. The Chern insulator near $\nu = 1$ shows vanishing $R_{xx}$ and quantized $R_{xy}$. Both $R_{xx}$ and $R_{xy}$ exhibit a weak dependence on $V_{CG}$. In the ideal case of a top graphite gate with perfect screening, $R_{xx}$ and $R_{xy}$ should be independent of $V_{CG}$. However, the few-layer graphite top gate has a relatively small quantum capacitance, allowing electric field from the control gate to weakly penetrate through and therefore inducing a weak gating effect on the mRG channel from $V_{CG}$.

The dependence of $R_D$ on $V_{CG}$ and $\nu$ for $V_{CG} \gtrsim -4$V is correlated with that for $R_{xx}$ and $R_{xy}$. The data supports the claim that the filling factor of the bulk mRG channel is nearly identical to that of the QPC constriction (at least for $V_{CG} \gtrsim -4$V). In Extended Data Fig. 3c, $R_D$ is nearly quantized at $\frac{h}{e^2}$ near $\nu = 1$ for $-1\text{V} \gtrsim V_{CG} \gtrsim -4$V, consistent with the weak tunneling regime we examined in the main text. For $V_{CG} \gtrsim -1$V, both the antidot and tunneling electrodes are increasingly electron-doped (note that $V_{BG}$ is kept at a positive voltage), leading to an enhanced tunneling probability between the two, and therefore a higher $R_D$ in this region. In the strong tunneling limit, $R_D$ would approach $\frac{2h}{e^2}$ due to electrical shunting through the antidot and tunneling electrodes.

For $V_{CG} \lesssim -4$V, the antidot and tunneling electrodes become hole-doped (while the bulk mRG channel remains electron-doped), forming a p-i-n junction at the antidot and an effective double dot structure: a p-type dot at the center and a n-type dot on the edge surrounding the p-type dot (Extended Data Fig. 4). The p-type regions in the device expands under a more negative $V_{CG}$ and would ultimately push the QPC constriction to near charge neutrality. In this strong tunneling limit, the edge states from the antidot and tunneling electrodes are hybridized; the device is effectively divided into two largely disconnected portions by the antidot and tunneling electrodes, leading to a diverging $R_D$. The strongly suppressed source-drain current in this regime also accounts for the increased noise in $R_{xx}$ and $R_{xy}$ observed in Extended Data Fig. 3a,b at large negative $V_{CG}$.

**Antidot in the p-i-n regime**

In the main text, we focus on the regime where the antidot has the same charge carrier type as the mRG bulk channel (i.e. n-n regime); here we discuss the situation where the antidot is p-doped under a large negative $V_{CG}$, forming a p-i-n junction. This scenario is illustrated in Extended Data Fig. 4a, where the dark blue regions are p-type and the red edge channels are n-type (the light blue bulk is incompressible). A double quantum dot structure is formed with the p-type dot at the center isolated from the n-type edge dot by an intrinsic (i) region. The two dots are capacitively coupled by Coulomb interactions.

Whereas the edge dot is surrounded by a bulk Chern insulator and thus features Laughlin charge pumping, the center dot is enclosed by a topologically trivial insulator (i.e. the i region) and no Laughlin charge pumping is expected. This distinction is shown in the measurements of $R_D$ as a function of $V_{CG}$ and $B$ in Extended Data Fig. 4b. The diagonal stripes, a feature of Laughlin charge pumping as discussed in the main text, come from the edge dot. In contrast to Fig. 2e and 3d, however, a weaker set of vertical stripes originated from the center dot is also observed (they are vertical because a magnetic field cannot pump charges into the center dot). These vertical stripes introduce regular shifts in the diagonal stripes whenever the two cross each other. The center dot features are weaker in tunneling measurements because tunneling occurs only between the edge dot and the tunneling electrodes; the center dot features are manifested only through its capacitive coupling to the edge dot. The two sets of stripes are schematically illustrated in Extended Data Fig. 4c. The charging periods are $\Delta V_{CG,edge} \approx 22$mV and $\Delta V_{CG,center} \approx 21$mV for the edge and center dot, respectively. They have similar magnitude; the value is about twice of $\Delta V_{CG} \approx 9.5$mV for the n-n dot investigated in the main text. The result shows that the edge and center dots in the p-i-n regime as well as the dot in the n-n regime all have similar gate capacitances.

**Electrostatics simulation for the antidot**

Due to the finite size of the antidot (about 210nm diameter) and the non-negligible vertical separation between the control gate and the mRG (about 41nm), a simple parallel-plate capacitor model is insufficient to accurately describe the electrostatics of the system. To this end, we performed quantitative electrostatics simulations using COMSOL with realistic device parameters.

We considered a four-layer structure consisting of the control gate (CG), top gate (TG), mRG, and bottom gate (BG) in the simulation. For simplicity, all conducting layers were modeled as metallic sheets. The top gate contains a circular hole with a diameter of 210nm, corresponding to the lithographically defined antidot. The hBN thicknesses between adjacent layers were determined by AFM measurements and independently verified by Shubnikov–de Haas oscillations, yielding $d_{CG} \approx 18$nm, $d_{TG} \approx 23$nm, and $d_{BG} \approx 19$nm from top to bottom, respectively, as illustrated in Fig. 1a. We used a relative dielectric constant of $\epsilon_{hBN} = 3$ for hBN.

In the simulation, we varied $V_{CG}$ while keeping $V_{TG} = V_{BG} = 0$V. The simulated potential profile near the antidot is shown in Extended Data Fig. 7a, where the different metallic layers are indicated by yellow lines. Extended Data Fig. 7b shows the corresponding charge distribution in the mRG layer. The lithographically defined antidot boundary is marked by vertical dashed lines. Near the antidot edges, the induced charge density is substantially reduced compared to the center region due to screening from the top gate.

Using the simulated charge distribution, we extracted a control-gate voltage $\Delta V_{CG} \approx 9.0mV$ required to add one electron to the antidot. This value is in good agreement with the experimentally measured oscillation periods of 9.5mV for $\nu = 1$ and 9.7mV for $\nu = 2/3$. In contrast, an ideal parallel-plate capacitor model would give $\Delta V_{CG} \approx 7.1$mV for the same antidot diameter, which is substantially smaller than both the experimental value and the realistic simulation result.

## Author Contributions

H.L., Z.G., C.X., Y.J., J.X., and L.M. fabricated the devices. H.L. performed the electrical measurements on devices 1-3 at Cornell. H.L., J.S., S.Y., and Z.H. performed the electrical measurements on device 3 at MIT. J.S. engineered the dilution fridge and electronic filters at MIT, and H.L., C.X., and Z.G. applied similar design and engineering to the dilution fridge at Cornell. K.W. and T.T. grew the bulk hBN crystals. H.L., J.S., and K.F.M. designed the scientific objectives and oversaw the project. All authors discussed the results and commented on the manuscript.

## Acknowledgement

We thank Mitali Banerjee and Xu Du for helpful advice on our experiment. We also thank Yuval Oreg and Allan MacDonald for helpful discussions.

## Figures

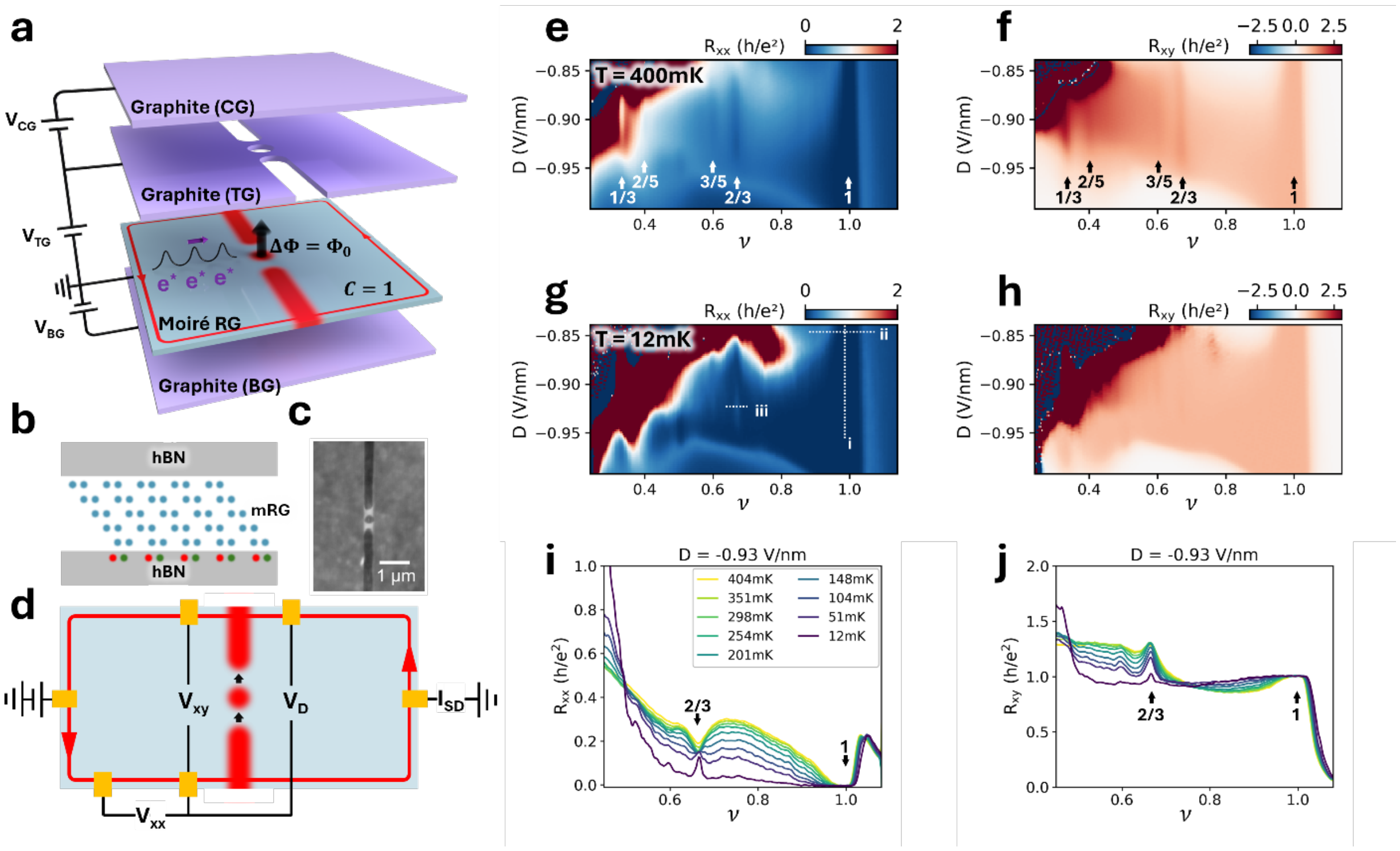

**Figure 1 | Gate-defined antidot in moiré rhombohedral graphene. a,b,** Schematic gate-defined antidot device (**a**) of pentalayer rhombohedral graphene (RG) angle-aligned with the bottom hBN to form a moiré lattice (**b**). CG, TG and BG denote the control gate, the lithographically patterned top gate and the bottom gate, respectively. $V_{CG}$, $V_{TG}$ and $V_{BG}$ are the voltages applied to CG, TG and BG, respectively. The moiré sample is set to an integer Chern insulating state with Chern number $C = 1$. Threading magnetic flux in units of the flux quantum $\Delta\Phi = \Phi_0$ pumps quasiparticles of charge $e^*$ to the antidot. **c,** An AFM image of the device near the antidot area. Scale bar denotes 1μm. **d,** Measurement schematics illustrating the source-drain current $I_{SD}$ and the longitudinal ($V_{xx}$), transverse ($V_{xy}$) and diagonal ($V_D$) voltage drops. The red-shaded areas denote the gate-defined antidot and tunneling electrodes; the red arrowed line illustrates the chiral edge state of a Chern insulator. **e-h,** Longitudinal ($R_{xx}$, **e** and **g**) and Hall ($R_{xy}$, **f** and **h**) resistances (in units of $\frac{h}{e^2}$) as a function of the filling factor ($\nu$) and the vertical displacement field ($D$) at temperature 400mK (**e** and **f**) and 12mK (**g** and **h**) and magnetic field 50mT. The arrows in **e** and **f** highlight the integer and fractional Chern insulating states. The dashed lines in (**g**) show the line cuts explored in Fig. 2 and 3. **i,j,** $R_{xx}$ (**i**) and $R_{xy}$ (**j**) versus $\nu$ at varying temperatures and at $D = -0.93$V/nm.

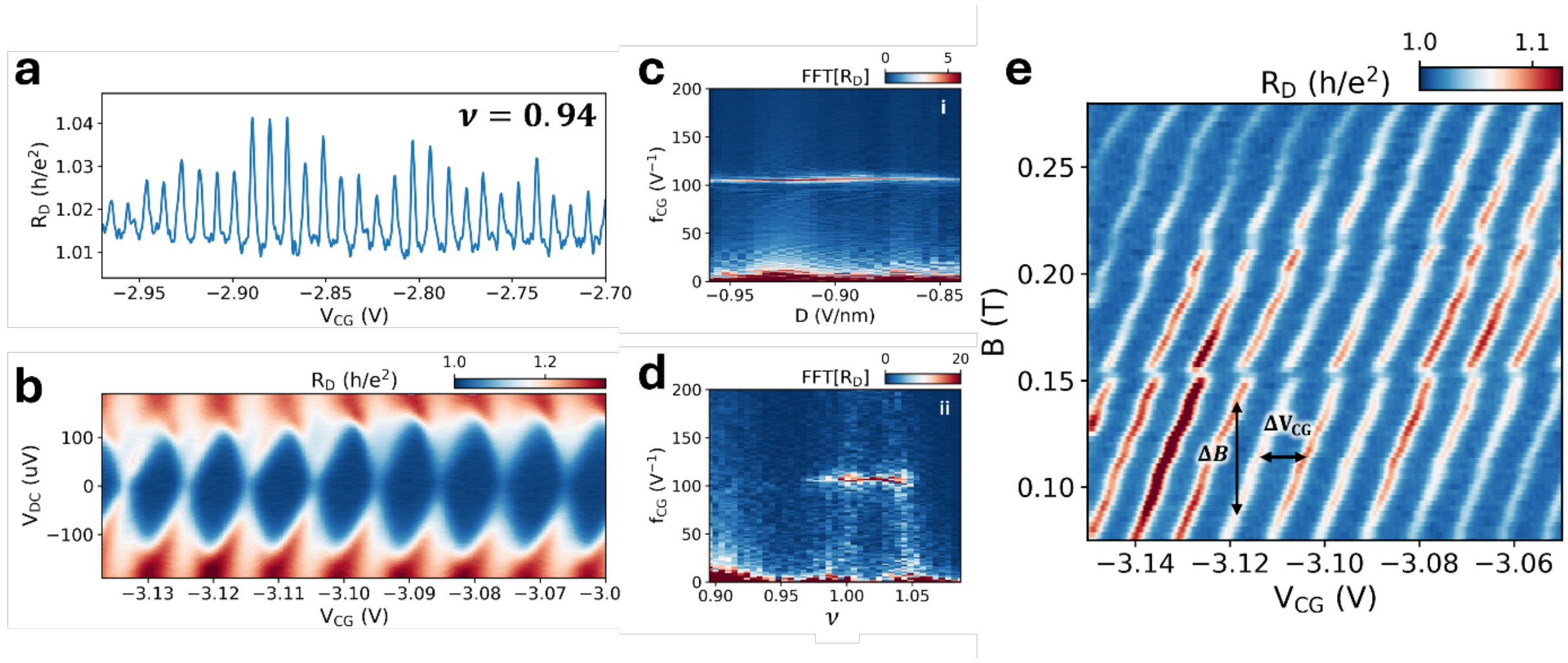

**Figure 2 | Antidot resonance tunneling near $\nu = 1$. a,** The diagonal resistance $R_D$ (in units of $\frac{h}{e^2}$) as a function of $V_{CG}$ measured at $\nu = 0.94$ and $D = -0.94$V/nm ($B = 50$mT). Each peak corresponds to a resonant quasiparticle tunneling event to the antidot. **b,** $R_D$ as a function of $V_{CG}$ and $V_{DC}$ at $\nu = 0.98$ and $D = -0.95$V/nm showing Coulomb diamonds. **c,d,** Fourier transformed spectrum of $R_D$ (with respect to $V_{CG}$) measured along the dashed lines in Fig. 1g: i ($\nu - 0.98$, **c**) and ii ($D = -0.85$V/nm, **d**). $f_{CG}$ denotes the Fourier transformed frequency of $V_{CG}$. **e,** $R_D$ as a function of $V_{CG}$ and $B$ at $\nu = 0.98$ and $D = -0.95$V/nm. The arrows denote the field ($\Delta B$) and the voltage ($\Delta V_{CG}$) periods for the Coulomb oscillations. All measurements were acquired at $T =$ 12mK.

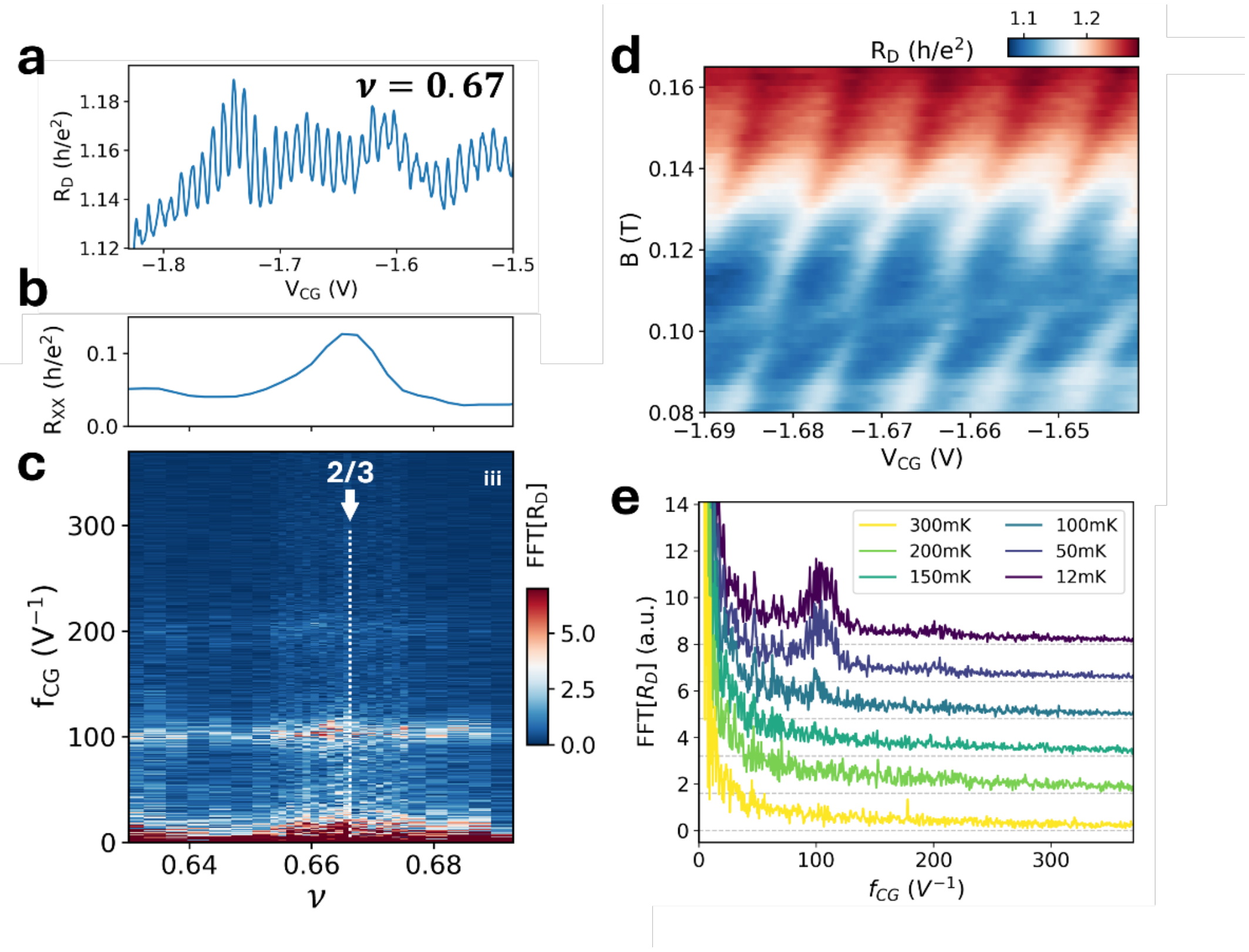

**Figure 3 | Antidot resonant tunneling near $\nu = 2/3$. a,** $R_D$ as a function of $V_{CG}$ measured at $\nu = 0.67$ ($B = 50$mT). **b,** $R_{xx}$ line cut along the dashed line iii in Fig. 1g. **c,** Fourier transformed spectrum of $R_D$ (with respect to $V_{CG}$) measured along the same dashed line as in **b**. The oscillation at $f_{CG} \approx 100$V$^{-1}$ peaks around $\nu = 2/3$. **d,** $R_D$ as a function of $V_{CG}$ and $B$ at $\nu = 0.67$. **e,** The Fourier transformed spectrum of $R_D$ at $\nu = 0.67$ and different temperatures. The curves are vertically displaced for clarity. The oscillation at $f_{CG} \approx 100$V$^{-1}$ disappears at about 100mK. All measurements were acquired at $D = -0.93$V/nm and $T = 12$mK (except panel **e** for temperature).

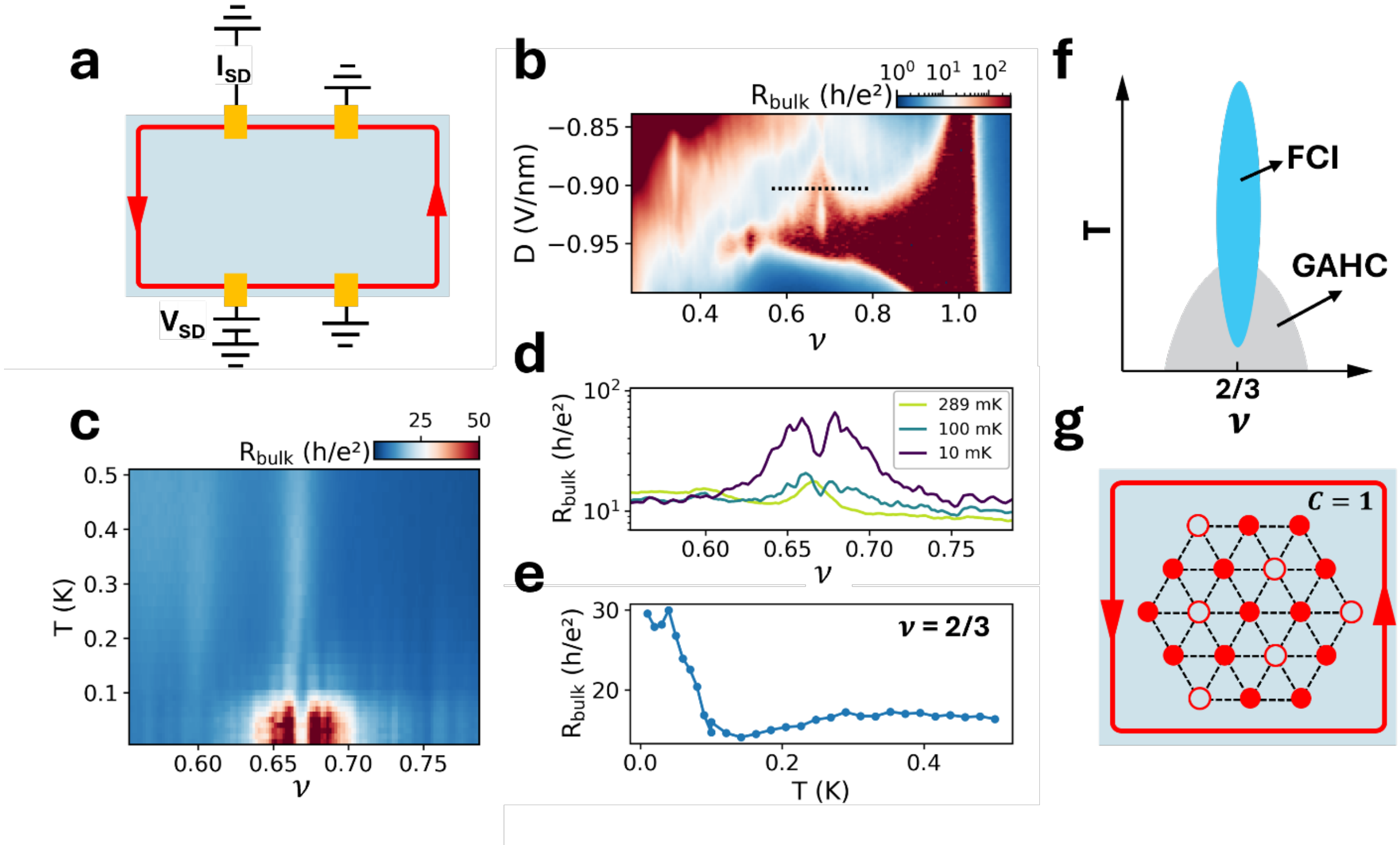

**Figure 4 | Competing ground states probed by bulk resistance. a,** Measurement schematics for the two-terminal quasi-bulk resistance $R_{bulk}$ of a Chern insulator. All electrodes downstream the source electrode (biased at $V_{SD}$) are grounded; currents are collected only at the electrode immediately upstream the source electrode. **b.** $R_{bulk}$ (in units of $\frac{h}{e^2}$) as a function of $\nu$ and $D$ at $T = 12\text{mK}$. **c,** $R_{bulk}$ as a function of $\nu$ and $T$ measured at $D = -0.90\text{V/nm}$ (i.e. along the dashed line in **b**). **d,e,** Selected line cuts from **c** at representative temperatures (**d**) and at $\nu = 2/3$ (**e**). **f,** Schematic phase diagram near $\nu = 2/3$ showing a FCI (stable at high temperatures) and a generalized anomalous Hall crystal (GAHC, stable at low temperatures). The two states compete near $\nu = 2/3$. **g,** Schematic structure of the GAHC carrying a chiral edge state and a generalized Wigner crystal in the bulk. Solid and empty dots represent electrons and holes, respectively. All measurements were acquired at $B = 50\text{mT}$.

## Extended Data Figures

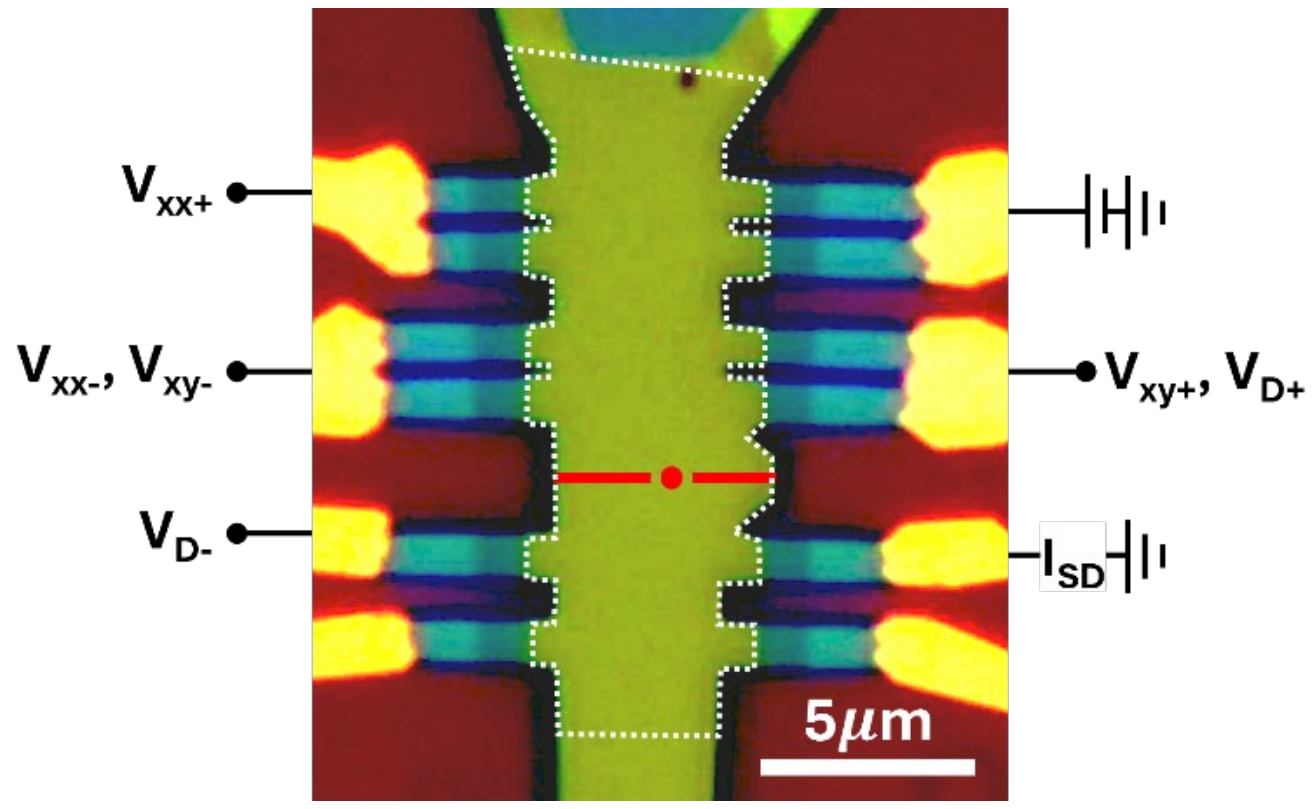

**Extended Data Figure 1 | Optical image for mRG antidot device.** Dashed lines enclose the mRG device channel. The AFM-patterned antidot and tunneling electrodes are marked in red. Measurement connections illustrating the source and drain contacts, and the voltage probes for $V_{xx} = V_{xx+} - V_{xx-}$, $V_{xy} = V_{xy+} - V_{xy-}$ and $V_D = V_{D+} - V_{D-}$ are shown. Scale bar is 5μm.

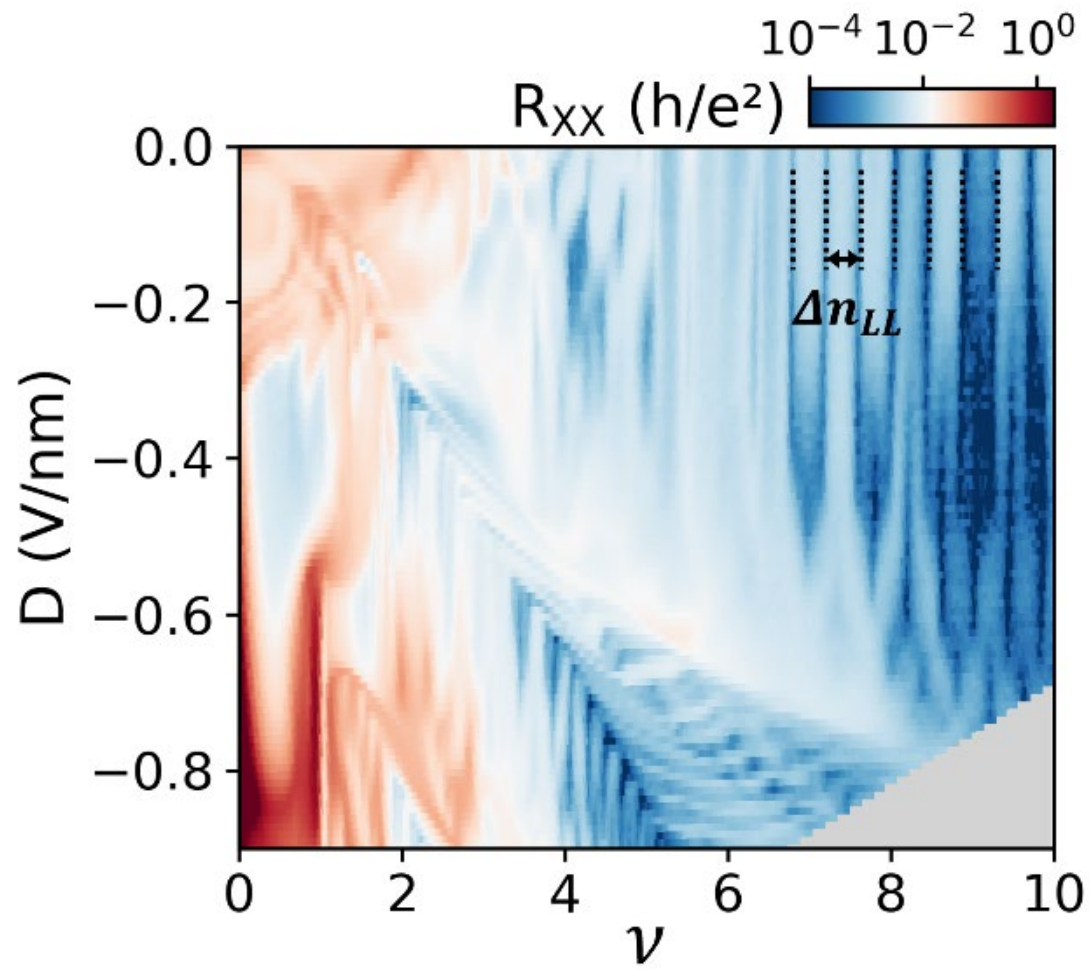

**Extended Data Figure 2 | Shubnikov-de Haas oscillations at $B = 3$T.** $R_{xx}$ as a function of $\nu$ and $D$ at $T = 12$mK and $B = 3$T. The moiré density $n_M$ was calibrated by the Landau level separations. The adjacent $R_{xx}$ minima (marked by dashed lines in the top-right corner of the map) correspond to filling of the successive four-fold-degenerate Landau levels in the full metal regime of rhombohedral graphene[64]. The density spacing $\Delta n_{LL} = 4eB/h = 0.29 \times 10^{12}\text{cm}^{-2}$ at $B = 3$T is equal to the filling factor spacing $\Delta\nu_{LL} = 0.40$ between adjacent minima. The moiré density is $n_M = \Delta n_{LL}/\Delta\nu_{LL} = 0.73 \times 10^{12}\text{cm}^{-2}$.

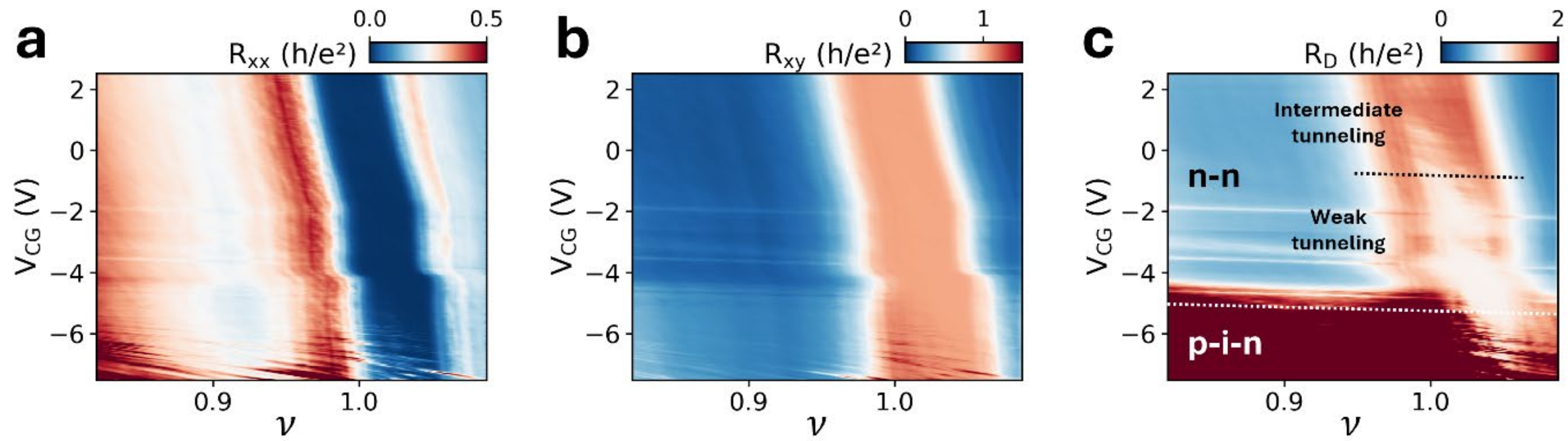

**Extended Data Figure 3 | Coarse control by $\boldsymbol{V_{CG}}$. a-c,** $R_{xx}$ (**a**), $R_{xy}$ (**b**) and $R_D$ (**c**) as a function of $\nu$ and $V_{CG}$ at $D = -0.83$V/nm ($T = 12$mK and $B = 50$mT), focusing on the $\nu = 1$ Chern insulator. Dashed lines in **c** separate the plots into three regimes: 1) Intermediate tunneling regime for a n-n antidot, 2) Weak tunneling regime for a n-n antidot, and 3) A p-i-n antidot. See Fig. 1d and Extended Data Fig. 4a, respectively, for schematics of a n-n and p-i-n antidot. Note that no Coulomb oscillations can be resolved in this coarse-range scan of $V_{CG}$. See Methods for detailed discussions.

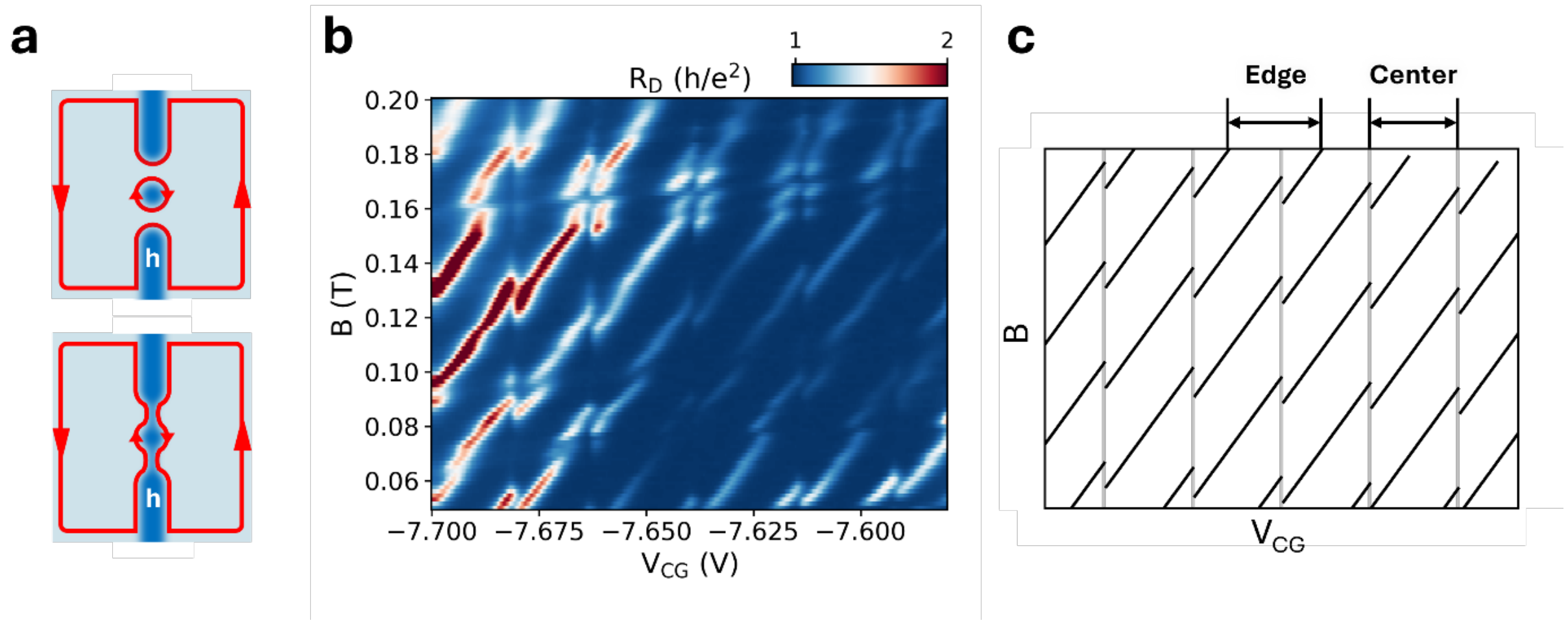

**Extended Data Figure 4 | Antidot charging in the p-i-n regime. a,** Schematic antidot device in the p-i-n regime. The center of the antidot is hole-doped (dark blue) and isolated from the n-type chiral edge state (red) by an intrinsic region of a topologically trivial insulator. The light blue region is incompressible. Top: Weak tunneling regime. The p-type center dot (dark blue) and the surrounding n-type edge dot (red) constitute a double quantum dot system. Bottom: Strong tunneling regime. The chiral edge states from the antidot and tunneling electrodes are hybridized, effectively separating the device into two halves. The QPC constriction is near charge neutrality in this regime. **b,** $R_D$ as a function of $V_{CG}$ and $B$ at $\nu = 0.98$ and $D = -0.95$V/nm in the p-i-n regime ($T = 12$mK). **c,** Schematic quasiparticle charging in a double dot system. The edge dot exhibits diagonal stripes (black) corresponding to Laughlin pumping while the center dot shows vertical stripes (gray) without Laughlin pumping. The arrows mark the charging period for the edge and center dots.

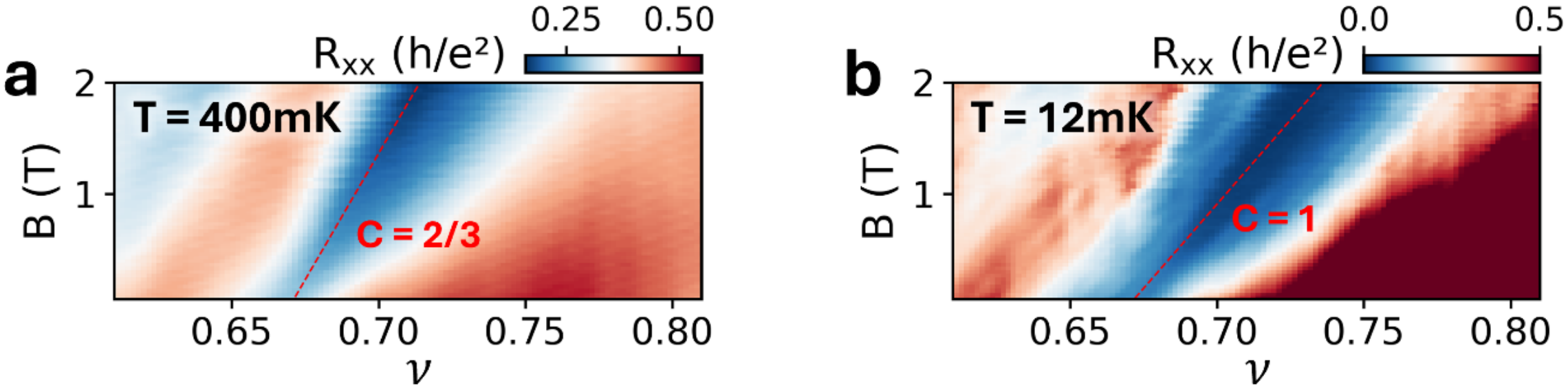

**Extended Data Figure 5 | Dispersion of the $\nu = 2/3$ Chern insulator under a magnetic field. a, b,** $R_{xx}$ as a function of $\nu$ and $B$ at (**a**) $T = 400$mK and (**b**) $T = 12$mK ($D = -0.88$V/nm). The slope $C$ can be extracted from the dispersion of the $R_{xx}$ minimum (red dashed lines) using the relation $n_M \cdot \frac{d\nu}{dB} = C \cdot \frac{e}{h}$. We obtain $C \approx 2/3$ at $T = 400$mK and $C \approx 1$ at $T = 12$mK, suggesting a thermodynamic phase transition from a FCI at high temperatures to an integer Chern insulator at low temperatures.

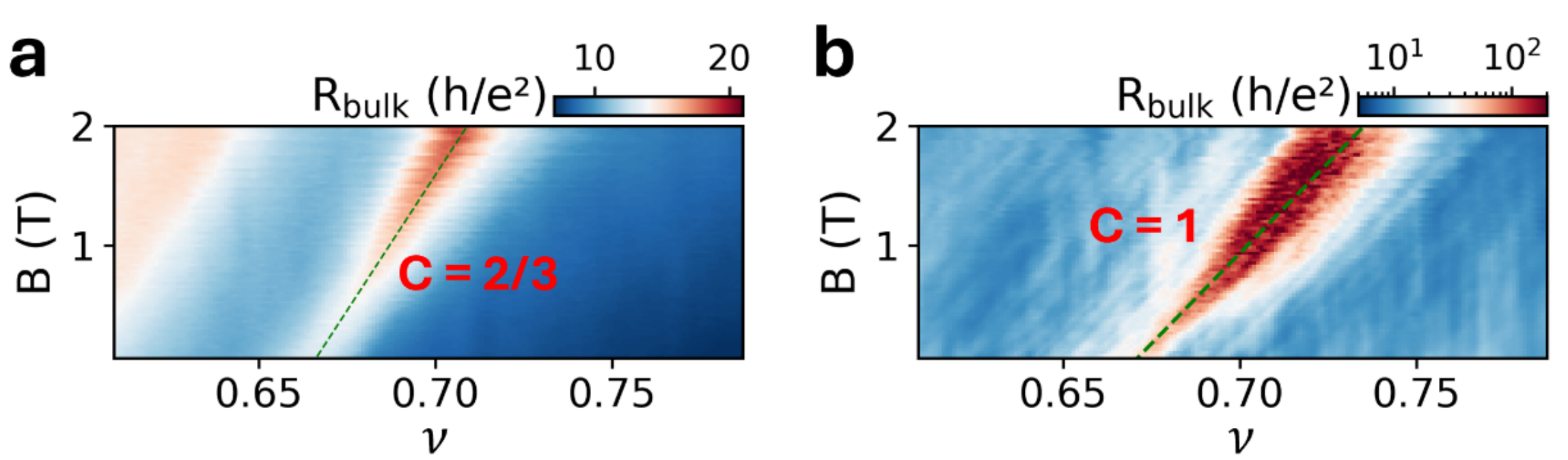

**Extended Data Figure 6 | Dispersion of the $\nu = 2/3$ Chern insulator from bulk resistance. a, b,** $R_{bulk}$ as a function of $\nu$ and $B$ at (**a**) $T = 400$mK and (**b**) $T = 12$mK ($D = -0.88$V/nm). The slope $C$ can be extracted from the dispersion of the $R_{bulk}$ peak (green dashed lines). Similar to Extended Data Fig. 5, we obtain $C \approx 2/3$ at $T = 400$mK and $C \approx 1$ at $T = 12$mK.

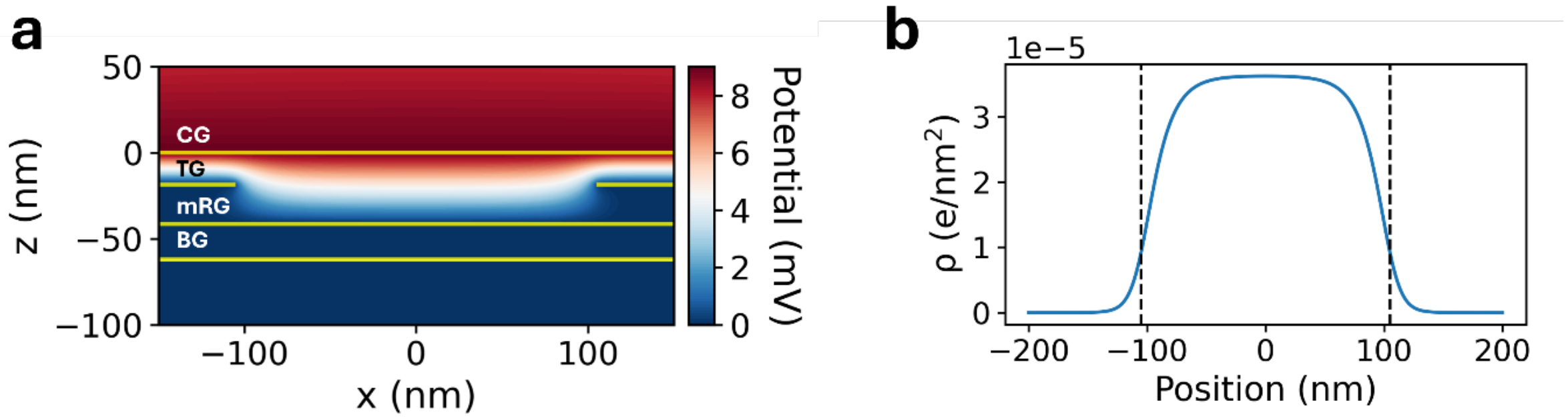

**Extended Data Figure 7 | Electrostatics simulations for the antidot. a,** Cross-sectional view of the simulated potential profile. The yellow lines mark the CG, TG, mRG and BG layers. **b,** Simulated charge density distribution through the center of the antidot in the mRG layer. The lithographically defined antidot boundaries are marked by vertical dashed lines. The simulation was performed at $V_{CG} = 9.0$mV and $V_{TG} = V_{BG} = 0$.

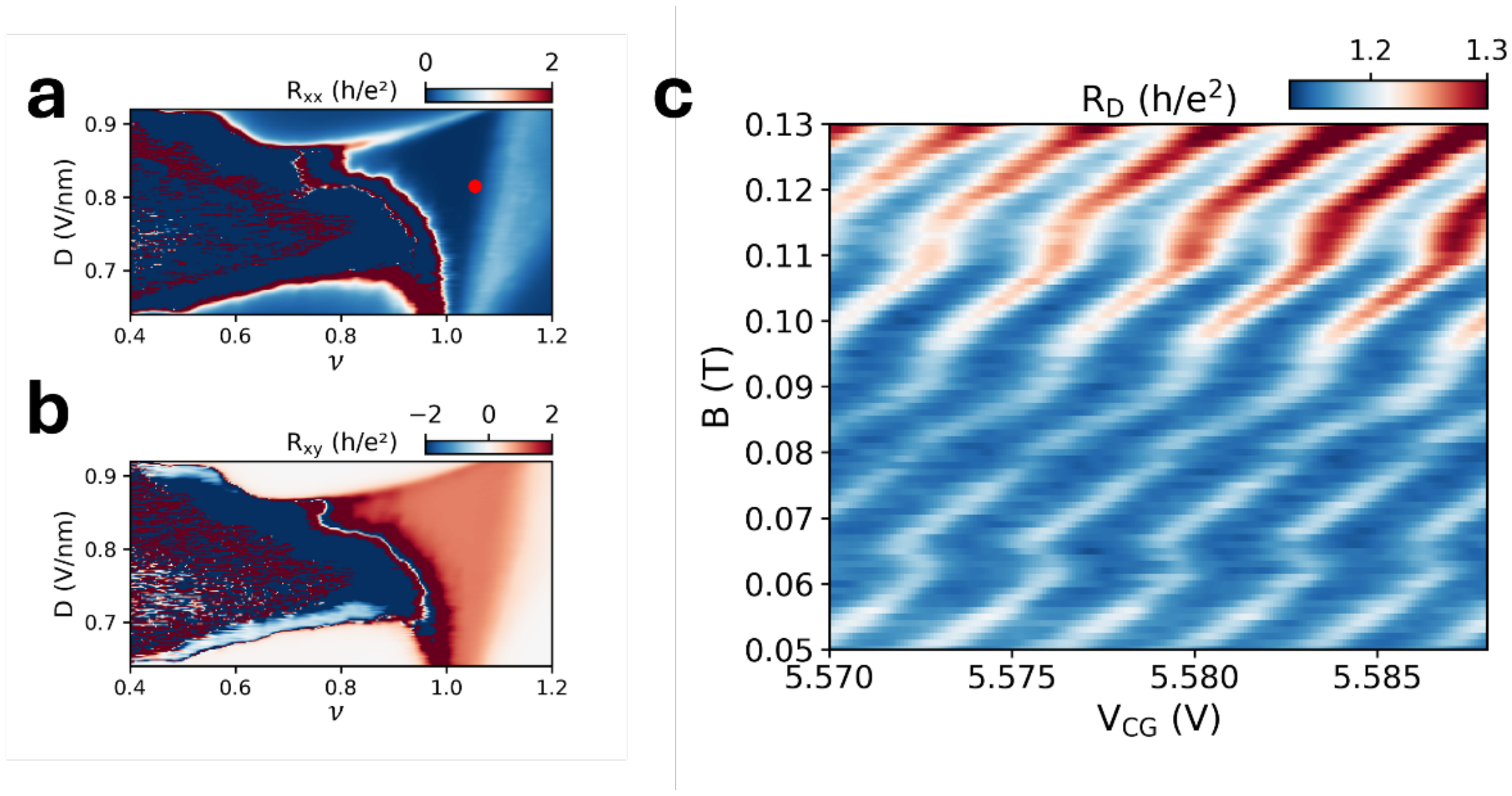

**Extended Data Figure 8 | Selected data from mRG antidot device 2. a,b,** Longitudinal ($R_{xx}$, **a**) and Hall ($R_{xy}$, **b**) resistances (in units of $\frac{h}{e^2}$) as a function of $\nu$ and $D$ under a magnetic field 50mT. Compared to the device in the main text, the rhombohedral graphene here is angle-aligned to the top hBN layer; the moiré period is 13.2nm and the moiré density is $0.66 \times 10^{12}$cm$^{-2}$. The lithographic antidot diameter is 310nm and the lithographic width of the QPC is 170nm. **c**, $R_D$ as a function of $V_{CG}$ and $B$ at $\nu = 1.06$ and $D = 0.83$ V/nm (red point in **a**). Both Coulomb oscillations and Laughlin charge pumping are observed. All measurements were acquired at $T =$ 12mK.

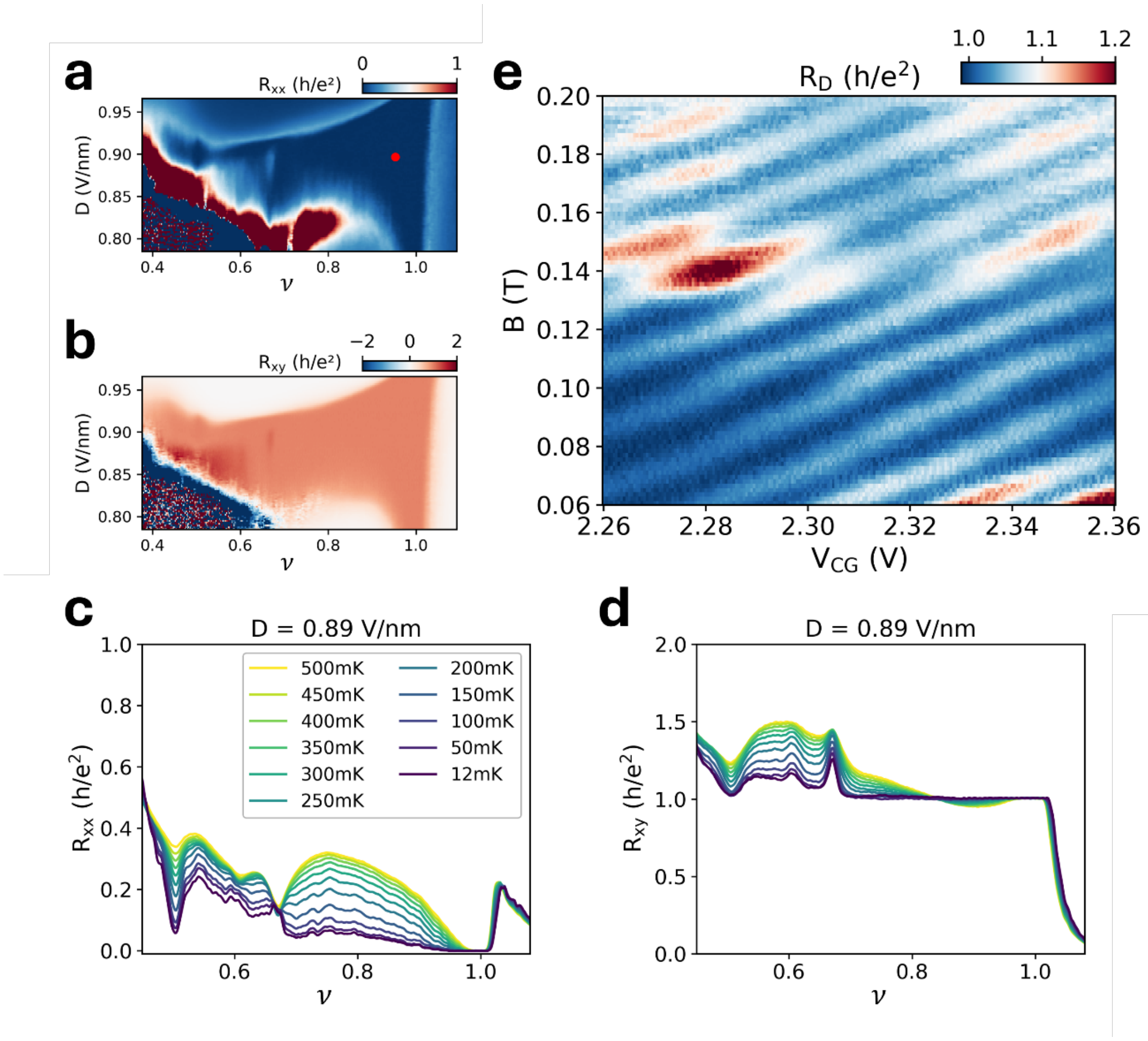

**Extended Data Figure 9 | Selected data from mRG antidot device 3. a,b,** Longitudinal ($R_{xx}$, **a**) and Hall ($R_{xy}$, **b**) resistances (in units of $\frac{h}{e^2}$) as a function of $\nu$ and $D$ under a magnetic field 50mT. Compared to the device in the main text, the rhombohedral graphene here is angle-aligned to the top hBN layer; the moiré period is 13.1nm and the moiré density is $0.67 \times 10^{12}$cm$^{-2}$. The lithographic antidot diameter is 265nm and the lithographic width of the QPC is 175nm. **c,d,** $R_{xx}$ (**c**) and $R_{xy}$ (**d**) versus $\nu$ at varying temperatures and at $D = 0.89$V/nm. **e**, $R_D$ as a function of $V_{CG}$ and $B$ at $\nu = 0.95$ and $D = 0.89$V/nm (red point in **a**). Both Coulomb oscillations and Laughlin charge pumping are observed. All measurements were acquired at $T = 7$mK unless otherwise specified.